\documentclass{article}
\usepackage{spconf,amsmath,graphicx}
\usepackage{booktabs}

\title{How Similar or Different Is Rakugo Speech Synthesizer to \\ Professional Performers?}
\name{Shuhei Kato$^{\star\dagger}$\sthanks{The first author is currently with RevComm Inc., Japan. This work was partially supported by JST, CREST, Japan, under Grant JPMJCR18A6, VoicePersonae Project, and AIP Challenge; and by MEXT KAKENHI, Japan, under Grant 16H06302, 18H04120, 18H04112, and 18KT0051.}, Yusuke Yasuda$^{\star\dagger}$, Xin Wang$^\star$, Erica Cooper$^\star$, Junichi Yamagishi$^\star$}
\address{$^\star$National Institute of Informatics, Japan
$^\dagger$The Graduate University for Advanced Sciences, Japan
}
\begin{document}
%
\maketitle
\begin{abstract}
We have been working on speech synthesis for \textit{rakugo} (a traditional Japanese form of verbal entertainment similar to one-person stand-up comedy) toward speech synthesis that authentically entertains audiences. In this paper, we propose a novel evaluation methodology using synthesized rakugo speech and real rakugo speech uttered by professional performers of three different ranks. The naturalness of the synthesized speech was comparable to that of the human speech, but the synthesized speech entertained listeners less than the performers of any rank. However, we obtained some interesting insights into challenges to be solved in order to achieve a truly entertaining rakugo synthesizer. For example, naturalness was not the most important factor, even though it has generally been emphasized as the most important point to be evaluated in the conventional speech synthesis field. More important factors were the understandability of the content and distinguishability of the characters in the rakugo story, both of which the synthesized rakugo speech was relatively inferior at as compared with the professional performers. We also found that fundamental frequency ($f_{o}$) modeling should be further improved to better entertain audiences. These results show important steps to reaching authentically entertaining speech synthesis.
\end{abstract}

\vspace{-2mm}
\section{Introduction}
\label{sec:intro}
\vspace{-2mm}

Entertainment has been essential to human beings since ancient times. While the role of the performer, a person who entertains people by acting, singing, dancing, or playing music~\cite{McIntoshC2013}, has been carried out by human beings almost exclusively for a long time, machines are beginning to carry out the role today, e.g., singing voice synthesizers~\cite{Yamaha2004,CeVIO2012,NEUTRINO2020} and dancing robots~\cite{Sharp2016,XINGO2020}. In the field of verbal entertainment, we have focused on \textit{rakugo}, which is a traditional Japanese form of verbal entertainment similar to one-person stand-up comedy and is popular even today, and we have developed rakugo speech synthesizers based on the Tacotron framework~\cite{KatoS2019rakugo,KatoS2020}. One open question is how to evaluate rakugo speech synthesizers. We can easily imagine that the standard listening test for naturalness is not sufficient for this purpose. Are there any good ways to benchmark their performance? 

In Japan, professional rakugo performers are ranked at one of three levels\footnote{Strictly speaking, \textit{Edo} (Tokyo) rakugo, which we focus on in this paper, has a class system, while \textit{Kamigata} rakugo, which was developed in Osaka and Kyoto, does not today.}, i.e., \textit{zenza} (minor performer), \textit{futatsume} (second-rank performer), and \textit{shin-uchi} (first-rank performer). In our previous study~\cite{KatoS2020}, we used a shin-uchi performer's audio recordings only as a reference for assessing our synthesizer. However, such evaluation may not be ideal since the skills of rakugo performers also vary significantly. We should compare rakugo speech synthesizers with rakugo performers from the three different ranks. This should clarify what is missing in our rakugo synthesizer to entertain audiences more precisely.

In this paper, we therefore propose a novel subjective evaluation methodology using natural speech uttered by performers from the three different ranks in addition to synthesized speech and show benchmarking results for our rakugo speech synthesizer. For this purpose, we recorded speech of a common story performed by a performer of each of the three ranks and then conducted a subjective comparison with synthesized speech of the same story in terms of five aspects: 1) naturalness, 2) distinguishability of characters in the story, 3) understandability of content, 4) degree of entertainment, and 5) performer's skill level. 

This paper is structured as follows. In Section 2, we explain the ranks of professional performers in rakugo. We describe audio data collected for this evaluation and our end-to-end TTS system in Sections 3 and 4, respectively. We then explain our evaluation methodology and its benchmark results in Section 5. Acoustic analysis results are shown in Section 6, and we summarize our findings in Section 7. 

\vspace{-2mm}
\section{Ranks of professional performers in Edo rakugo}
\vspace{-2mm}

Rakugo is divided into \textit{Edo} (Tokyo) rakugo and \textit{Kamigata} rakugo, which was developed in Osaka and Kyoto. In this paper, we focus on Edo rakugo. Edo rakugo has a three-rank class system. In ascending order, professional rakugo performers are ranked at either \textit{zenza}, \textit{futatsume}, or \textit{shin-uchi}. It generally takes three to five years to be promoted from zenza to futatsume and about ten years to be promoted from futatsume to shin-uchi. As of 2020, about 600 professional performers are active in Edo rakugo~\cite{RakugoKyokai2020,RakugoGeijutsuKyokai2020,TokyoKawaraban2018,RakugoTatekawaryu2020}.

\vspace{-2mm}
\section{Audio Recordings}
\label{sec:recording}
\vspace{-2mm}

How high is the skill level of our rakugo speech synthesizer compared with professional rakugo performers? To investigate this, we recorded the performances of a common story\footnote{Specific wording depends on performers because rakugo stories do not have any scripts.} told by professional performers of the three ranks.

The recordings were carried out in January 2020. The performers were Yanagiya Kogoto
(zenza), Ryutei Ichido (futatsume), and Yanagiya Sanza (shin-uchi, the same performer who performed for the database used in the speech synthesis model training~\cite{KatoS2020}). The recording conditions were the same as those of the recordings for the database. Each performer performed alone in a recording booth, and he did not face or receive any reactions from an audience.
The story performed by them is called ``Misomame.'' The total durations of the recordings by the zenza, futatsume, and shin-uchi were 2.5, 2.7, and 4.2 minutes, respectively. To record performances that sound as natural as possible, we did not re-record any of the stories when mispronunciations or restatements occurred, except in cases where the performer asked us to do so.\footnote{The shin-uchi performer attended the recording session of the zenza performer, and he supervised and instructed the zenza performer when necessary.}

\vspace{-2mm}
\section{Speech synthesis model}
\label{sec:speech_synthesis_model}
\vspace{-2mm}

We used a variant of the Tacotron-based TTS system (\textsf{SA-Tacotron-context} model from our previous study~\cite{KatoS2020}) because this model was evaluated as the best one. This model takes textual information and context embeddings as inputs. It should be noted that the model is based on speech by the shin-uchi performer recorded in 2017, and the newly recorded speech in Section \ref{sec:recording} was not used for model building and was used only for comparison with synthesized speech. Please refer to \cite{KatoS2020} for details on the features, network structure, training conditions, etc. Minor differences from \cite{KatoS2020} were as follows. 1) The sentences in ``Misomame'' were excluded from the training set and validation set. As a result, we used 6,362 sentences (3.67 hours) for training, 706 sentences (0.42 hours) for validation, and 273 sentences (0.22 hours) for testing. 2) The sampling frequency was changed from 16\,kHz to 24\,kHz for mel-spectrogram output from the speech synthesis model and waveforms generated through a WaveNet~\cite{OordA2016} vocoder~\cite{WangX2018,TamamoriA2017}. Accordingly, the frame shift and fast Fourier transform size were changed to 12\,ms and 2,048, respectively.

\vspace{-2mm}
\section{Listening test for benchmark}
\vspace{-2mm}

To benchmark the level of our rakugo speech synthesis, we designed a new listening test and conducted a large-scale listening test as described below. 

\vspace{-3mm}
\subsection{Test conditions}
\vspace{-1mm}

Speech of ``Misomame'' was used in the test, although speech of shorter stories was used for a listening test in our previous study. The reason is that we believe a ``full'' story\footnote{While there is no clear definition either of a full rakugo story or short story, short stories tend to appear in the makura, or prelude to the main story, of the performance of a full story, and are never independently performed on a stage. In Edo rakugo, several hundred traditional stories are performed.} is more suitable for evaluating the level of rakugo speech synthesis than a short story. We therefore adopted ``Misomame,'' which is a full story, though relatively short in duration, on the basis of advice from Yanagiya Sanza, the shin-uchi performer above.

The speech was synthesized sentence by sentence. Pauses between sentences were not predicted\footnote{They should be predicted, but that is out of the scope of this paper.}, and the pauses between sentences for the synthesized speech were the same as those of the real audio recording. Listeners evaluated the speech {\em not} sentence by sentence but as a whole story. All speech was normalized to $-26$\,dBov over the whole story using sv56~\cite{ITU2005}.

We asked listeners to answer a five-scale mean opinion score (MOS) based test. Listeners listened to either speech by the professional performers (zenza, futatsume, or shin-uchi) or the synthesized speech, and they evaluated them according to the five questions below.
\begin{description}
\setlength{\parskip}{0cm} 
\setlength{\itemsep}{0cm} 
\item[1)] How natural did the performer sound?
\item[2)] How accurately did you think you could distinguish each character?
\item[3)] How well did you think you could understand the content?
\item[4)] How well were you entertained?
\item[5)] How high was the rakugo skill level of the performer?
\end{description}
The most important question was Q4 since rakugo is a form of verbal entertainment. Q5 was intended for evaluating the ``skill level'' of the rakugo speech synthesis as if it were a professional performer. The others were questions about factors that we hypothesized may affect the results of Q4 and Q5. A total of 292 listeners participated in 292 evaluation rounds.

\vspace{-3mm}
\subsection{Results}
\vspace{-1mm}

\begin{figure*}[t]
    \begin{center}
    \vspace{-2mm}
    \includegraphics[width=2\columnwidth]{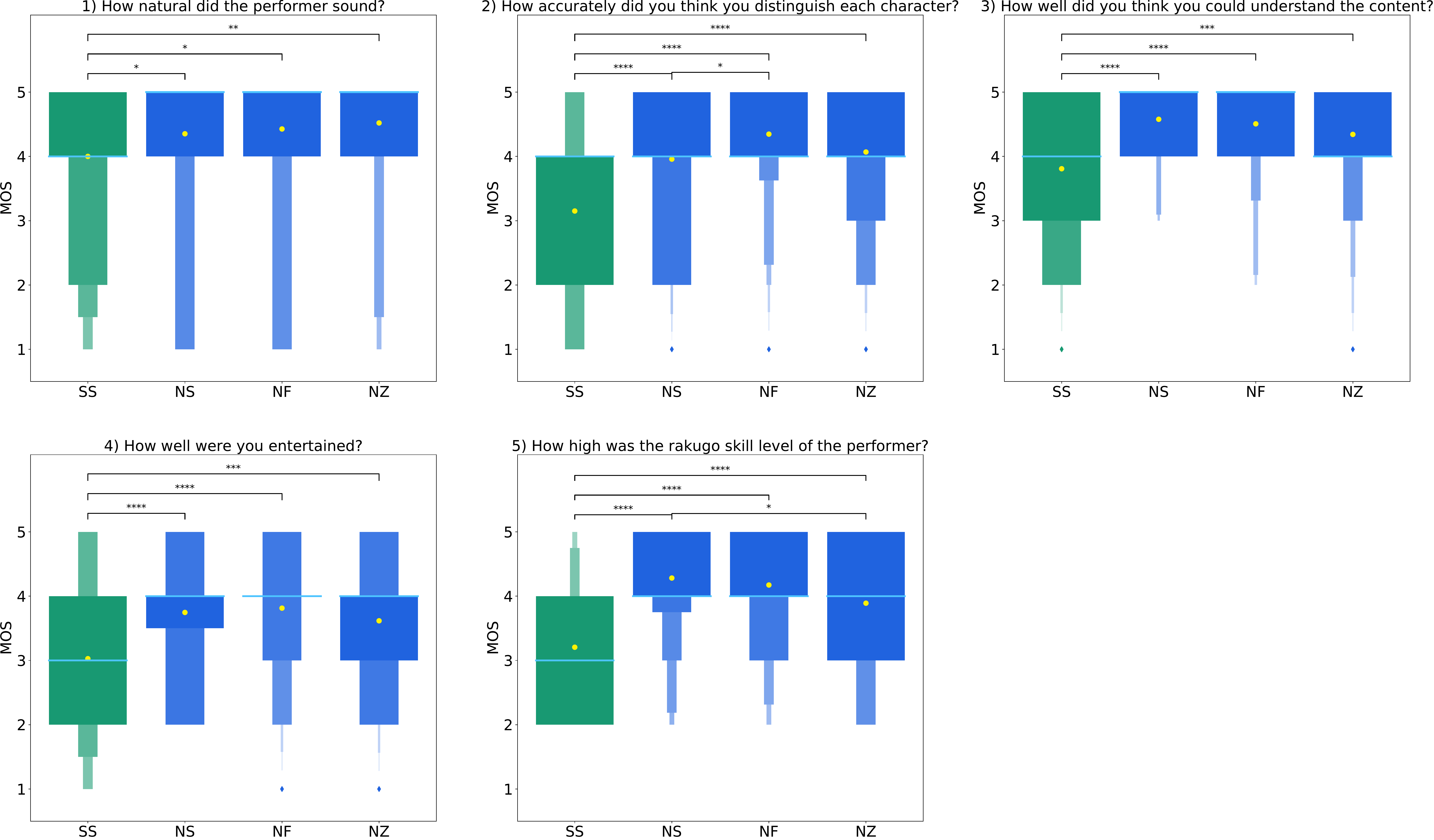}
    \vspace{-1mm}
    \caption{Boxen plots for each question of listening test. Light blue lines represent medians, and yellow dots represent means. *: $p < 0.05$, **: $p < 0.01$, ***: $p < 0.005$, ****: $p < 0.001$.}
    \label{fig:Q1_Q5}
    \end{center}
    \vspace{-6mm}
\end{figure*}

The listening test results are shown in Fig.~\ref{fig:Q1_Q5}, where \textsf{SS}, \textsf{NS}, \textsf{NF}, and \textsf{NZ} correspond to speech synthesis, shin-uchi, futatsume, and zenza, respectively. For statistical analysis, we used a Brunner-Munzel test~\cite{Brunner2000} with Bonferroni correction. 

As can be seen from the figure, the scores of the speech synthesis did not reach those of the natural speech of the professional performers, but we see that the trends in the score differences were different depending on the question.

For Q1 (naturalness), the mean score for the speech synthesis was 4.0. This means that the naturalness of the synthesized speech was high and comparable enough to that of natural speech. On the contrary, for Q2 (character), Q3 (content), and Q4 (entertaining), the mean scores for the speech synthesis ranged between 3.0 and 4.0, which were much lower than those for the professional performers. For Q3 and Q4, the $p$-values between the scores for the speech synthesis and those for the zenza were also smaller than the $p$-values between the scores for the speech synthesis and those for the futatsume or shin-uchi.

For Q5, which measures the skill level of the performer, the mean scores descended according to rank (shin-uchi $>$ futatsume $>$ zenza) as we expected. The synthesized speech was rated lower than the natural performances.

\vspace{-3mm}
\subsection{Correlations among questions}
\vspace{-1mm}

To understand the listening test results better, we calculated the correlation coefficients of the MOSs among the questions, and the results are shown in Table~\ref{tb:cc_MOS}. We see that the Q4 (entertaining) scores had a larger correlation coefficient in the order of the scores for Q5 (skill), Q3 (content), Q2 (character), and Q1. In other words, Q1 (naturalness) had the weakest correlation coefficient with Q4 (entertaining). The correlation coefficient between the scores for Q2 and those for Q3 was also relatively large. In summary, while the skill level (Q5), entertainment (Q4), understandability (Q3), and distinguishability of the characters (Q2) were correlated with each other to a moderate degree, naturalness (Q1) appeared to be less correlated with the other metrics. 

\begin{table}[t]
\small
\vspace{-2mm}
\begin{center}
\caption{Correlation coefficients of MOSs between questions.}
\label{tb:cc_MOS}
\vspace{1mm}
\begin{tabular}{lllll}
\toprule
 & Q2 & Q3 & Q4 & Q5 (skill)\\
\midrule
Q1 (naturalness) & 0.287 & 0.303 & 0.317 & 0.339\\
Q2 (character) & - & 0.538 & 0.486 & 0.580\\
Q3 (content) & - & - & 0.597 & 0.582\\
Q4 (entertaining) & - & - & - & 0.656\\
\bottomrule
\end{tabular}
\end{center}
\vspace{-7mm}
\end{table}

From the above results in Fig.~\ref{fig:Q1_Q5}, we learned that, even though the naturalness of the synthesized rakugo speech was close to that of the human professionals, it could not sufficiently entertain the listeners because the listeners could not perfectly distinguish characters in the synthesized speech and therefore could not adequately understand the content. In other words, we should not only improve the naturalness of synthesized speech but also refine the modeling of other aspects of speech, such as the distinguishability of characters in the case of rakugo, to better entertain listeners.

\vspace{-1mm}
\section{Acoustic analysis}
\vspace{-2mm}

\begin{figure}[t]
    \begin{center}
    \includegraphics[width=0.6\columnwidth]{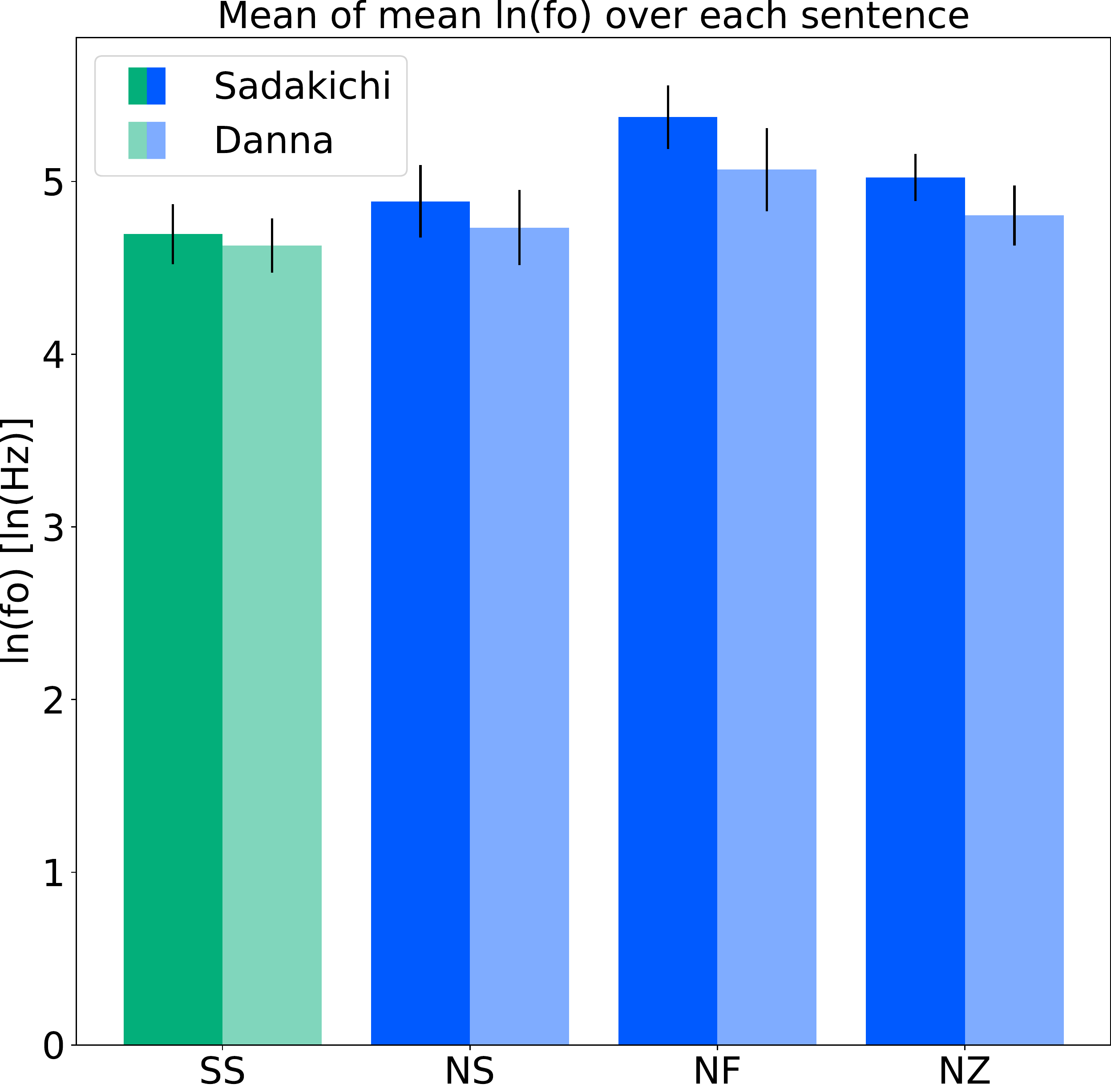}
    \vspace{-4mm}
    \caption{Means of means and standard deviations of logarithmic fundamental frequency (ln$f_{o}$) over each sentence. \textit{Sadakichi} and \textit{Danna} are two characters performed by performers.}
    \label{fig:mean_fo}
    \end{center}
    \vspace{-5mm}
\end{figure}

\begin{figure}[t]
    \begin{center}
    \includegraphics[width=0.9\columnwidth]{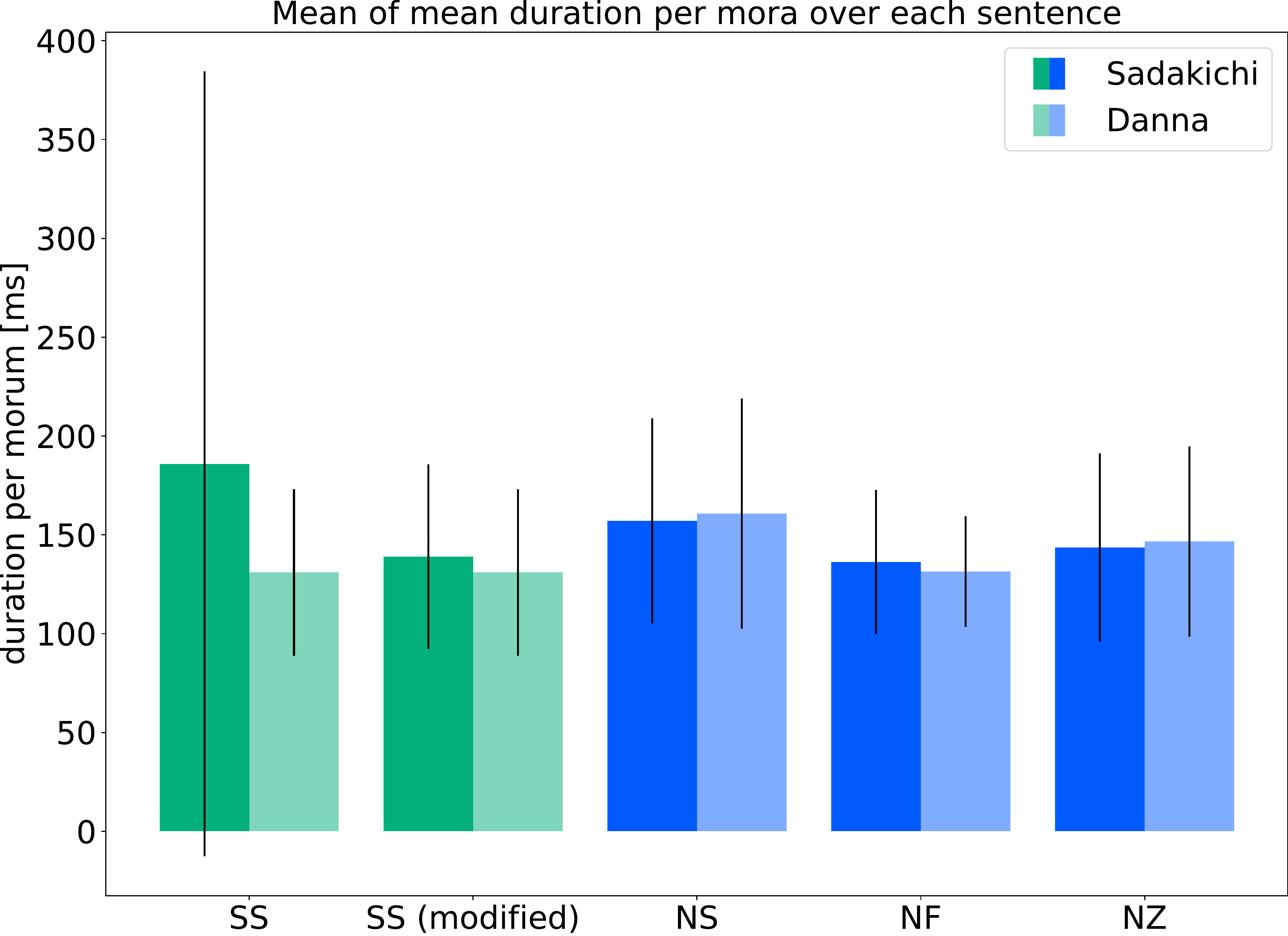}
    \vspace{-3mm}
    \caption{Means of means and standard deviations of duration per mora over each sentence. SS (modified) was calculated on basis of sentences, excluding two sentences for which duration was estimated as too long.}
    \label{fig:mean_dur}
    \end{center}
    \vspace{-6mm}
\end{figure}

We further investigated what makes it difficult for listeners to distinguish characters. We calculated the mean and standard deviation of the logarithmic $f_{0}$ (ln$f_{0}$) and duration per mora, sentence by sentence, and averaged them over the story for each character. Misomame has two characters, \textit{Sadakichi} (a boy) and \textit{Danna} (a middle-aged male). The results for the ln$f_{0}$ and duration corresponding to the two characters in the test set are shown in Figs. \ref{fig:mean_fo} and \ref{fig:mean_dur}. 

In Fig.~\ref{fig:mean_fo}, we can see that the cross-character difference for the mean ln$f_{0}$ of the synthesized speech was smaller than that of the human professionals' speech, particularly that of the futatsume's speech. We should consider that the extent to which a performer differentiates the voices of different characters depends on the performer. Yanagiya Sanza, the shin-uchi performer, does not strongly distinguish characters, according to an interview~\cite{TBA2011}. However, the difference for the synthesized speech was even smaller than that of the shin-uchi performer. We could therefore conclude that our speech synthesis does not have sufficient enough ability to distinguish characters using $f_{0}$.

In Fig.~\ref{fig:mean_dur}, we can see that all of the human professionals did not strongly distinguish characters using speech rates in the case of ``Misomame.'' This was the same for speech synthesis. We would like to note that professional performers may clearly change speech rates depending on the character~\cite{KatoS2018WaveNet}, and some synthesized speech samples used in our previous study~\cite{KatoS2020} distinguished characters using speech rates much more mildly than in the natural samples.

\begin{figure}[t!]
    \begin{center}
    \vspace{-3mm}
    \includegraphics[width=0.86\columnwidth]{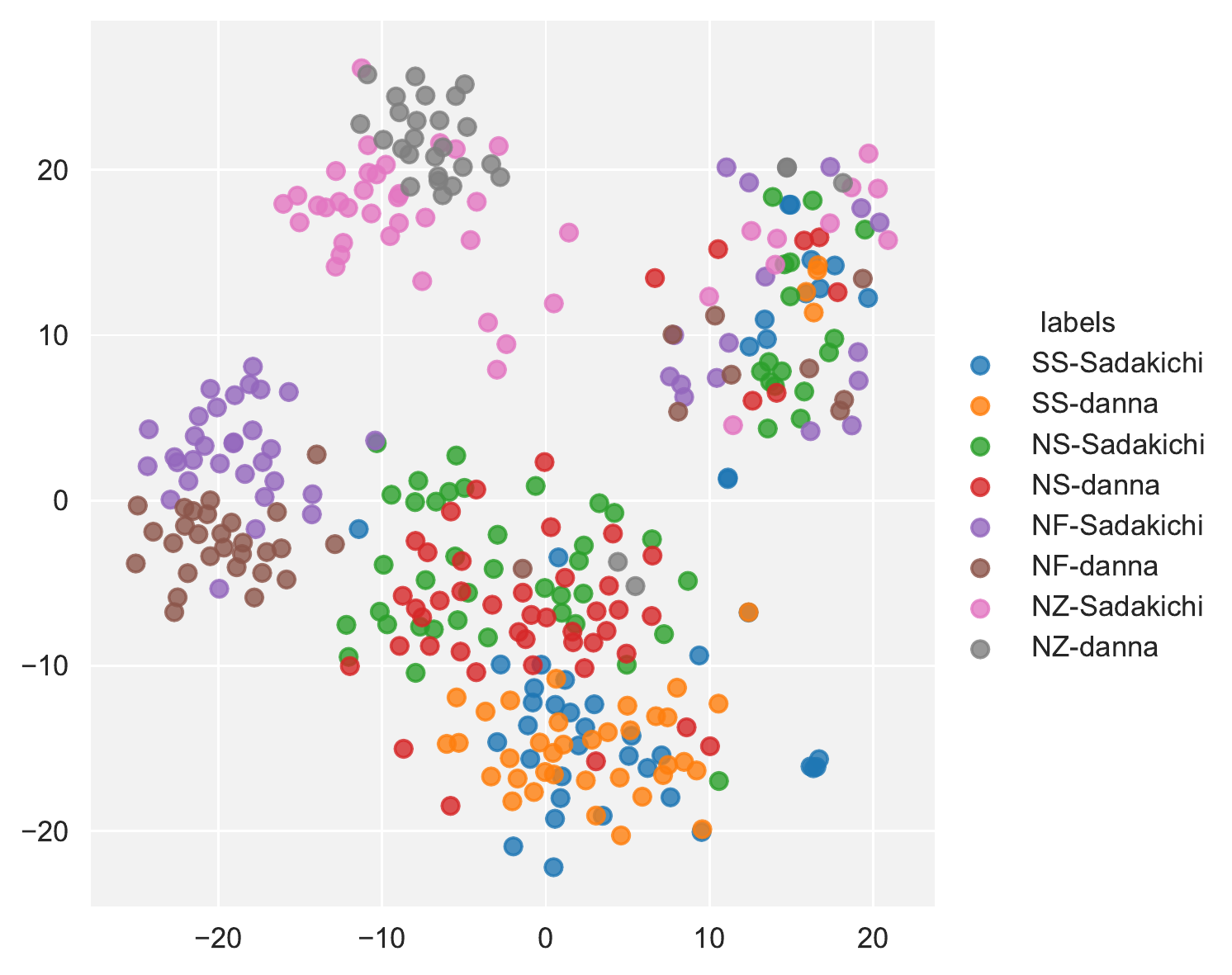}
    \vspace{-3mm}
    \caption{Visualization of x-vector for each sentence using t-SNE.}
    \label{fig:tsne}
    \end{center}
    \vspace{-6mm}
\end{figure}

How about speaker individuality? Fig.~\ref{fig:tsne} is a t-SNE~\cite{MaatenL2008} visualization of the x-vector~\cite{SnyderD2018} for each sentence. We can see four clusters, namely, ``zenza,'' ``futatsume,'' ``shin-uchi and speech synthesis,'' and ``all the systems.'' The ``zenza'' and ``futatsume'' clusters were generally divided into sub-clusters by character. In comparison, the ``shin-uchi'' cluster did not have clear separation by character, even though the listeners could distinguish the characters according to our listening test results. We therefore consider that the shin-uchi performer used different acoustic cues to express the characters from those of the lower rank performer. The ``speech synthesis'' cluster did not have separation by character, either. 

\vspace{-1mm}
\section{Conclusion}
\vspace{-2mm}

In this paper, we proposed a novel methodology for evaluating rakugo speech and conducted a listening test to investigate how the level of rakugo speech synthesis compares to professional rakugo performers at various levels. From the listening test results, we found that the level of speech synthesis did not reach that of human professionals. The results suggest, however, that we should make the $f_{0}$ expression of speech synthesis richer to better entertain audiences.

In future work, we will design a speech synthesis architecture and training framework that can better distinguish characters. The frequency of the properties of the characters (gender, age, social rank, etc.) in common rakugo stories, however, is very unbalanced. For example, young townsmen appear in rakugo stories very frequently, and women servants to samurai warriors rarely appear. We should consider such imbalance when designing a model. We will also work on other issues to be solved, such as estimating pauses between sentences and visual synthesis\footnote{Rakugo is essentially a form of audio-visual entertainment.}.



\normalsize

\vfill\pagebreak

\bibliographystyle{IEEEtrans}
\bibliography{mybib}

\end{document}